\documentclass[runningheads]{llncs}
\usepackage[T1]{fontenc}
\usepackage{graphicx}
\usepackage{hyperref}
\usepackage{amsmath}
\usepackage{wrapfig}
\usepackage{float}
\usepackage{amssymb}
\usepackage{geometry} 
\usepackage{placeins}

\title{SocraticAI: Transforming LLMs into Guided CS Tutors Through Scaffolded Interaction}
\titlerunning{SocraticAI: Scaffolded LLM Tutoring}
\author{Karthik Sunil \and Aalok Thakkar}
\authorrunning{K. Sunil \& A. Thakkar}
\institute{Ashoka University\\ 
\email{\{karthik.sunil\_ug25, thakkar\}@ashoka.edu.in}}

\begin{document}

\maketitle

\begin{abstract}
We present \textbf{SocraticAI}, a scaffolded AI tutoring system that integrates large language models (LLMs) into undergraduate Computer Science education through structured constraints rather than prohibition. The system enforces well-formulated questions, reflective engagement, and daily usage limits while providing Socratic dialogue scaffolds. Unlike traditional AI bans, our approach cultivates responsible and strategic AI interaction skills through technical guardrails including authentication, query validation, structured feedback, and RAG-based course grounding. Initial deployment demonstrates that students progress from vague help-seeking to sophisticated problem decomposition within 2--3 weeks, with over 75\% producing substantive reflections and displaying emergent patterns of deliberate, strategic AI use.
\end{abstract}

\section{Area and Context}
The rapid adoption of LLMs has transformed programming workflows but also introduced profound pedagogical challenges. In undergraduate CS education, unrestricted AI access often leads to shallow dependency and solution copying rather than genuine conceptual understanding \cite{becker2023generative}. Institutional responses have largely polarized between strict prohibition and unrestricted freedom, neither of which effectively balance innovation with academic integrity.
Prior research indicates that AI tutors can enhance learning when students remain actively engaged and reflective \cite{lyu2024effectiveness,liu2024teaching,sarsa2022automatic,liffiton2023codehelp,prather2023robots,macneil2022experiences,finnie2022robots}. However, persistent issues such as prompt flooding, superficial questioning, and uncritical copying, limit their pedagogical impact \cite{denny2023conversing,lau2023ban}. Contemporary frameworks increasingly argue that \emph{learning to interact productively with AI} is itself a critical component of digital literacy \cite{wm2024asking,ray2023chatgpt,kazemitabaar2023studying}.

\section{Best Practice}

We propose \textbf{SocraticAI}, which reimagines LLMs not as answer engines but as structured tutors within a guided, metacognitively informed learning framework. The system enforces constraints that require students to articulate their reasoning before receiving feedback, reflect afterward, and operate within deliberate-use limits. Key design components include:

\begin{enumerate}
\item \textbf{Constrained Interaction Design:} Daily query limits (8 per student) are enforced at the system level to discourage over-reliance and encourage deliberate question formulation. Students must submit structured input consisting of their current understanding, attempted solutions, or relevant code excerpts before receiving AI feedback. The input layer includes validation checks for completeness and relevance.

\item \textbf{Guided Query Framework:} To prevent direct solution-seeking, multi-stage prompting enforces a structured dialogue: 
\begin{enumerate}
    \item[(a)] describe the current approach and specific confusion points, 
    \item[(b)] explain prior attempts, and
    \item[(c)] identify the concept or implementation detail requiring clarification.
\end{enumerate}
Few-shot exemplars embedded in the system prompt illustrate productive Socratic dialogue patterns, reducing unproductive or overly general queries.  

\item \textbf{RAG-Based Course Integration:} Course materials are preprocessed into a semantically indexed knowledge base. A retrieval-augmented generation (RAG) pipeline grounds AI responses in relevant lectures, assignments, or textbook excerpts, minimizing hallucinations and ensuring curricular alignment. Students thus receive feedback contextualized within their course ecosystem.

\item \textbf{Reflection and Escalation:} Each interaction concludes with reflection prompts designed to elicit awareness (such as ``What did you learn?'' or ``What remains unclear?''). Students must summarize what they learned, identify unresolved questions, or outline next steps. These student interactions are systemically preserved for creating a record of student engagement. Additionally,  When reflection fails to resolve confusion, the system supports escalation to instructors, with full conversation history preserved for context-aware intervention. This record could further help instructors identify common learning challenges and tailor follow-up guidance accordingly.  

\item \textbf{System Architecture:} The system is implemented with modular, service-oriented architecture with separate services for authentication, feedback collection and handling, admin dashboard, vector retrieval, etc. This design choice facilitates scalable deployment and maintainable growth. Redis is used for real-time data storage with periodic persistence for post-analysis. The system also supports dynamic feedback tagging, and administrative monitoring via a dashboard. 

\item \textbf{Implementation and Observability:} A Prometheus-based observability pipeline logs query volume, reflection quality, and escalation frequency. Technical safeguards include (i) input sanitization to prevent injection attacks, (ii) context management for long conversations, and (iii) adversarial testing of system prompts. These guardrails ensure robustness, transparency, and resilience in real-world classroom settings.  
\end{enumerate}

Collectively, these mechanisms transform student use of LLMs from unstructured solution-seeking into a scaffolded process that cultivates problem decomposition, metacognitive reflection, and professional-grade AI interaction skills.

\section{Justification}
Our approach addresses three interrelated challenges in computer science education: fostering AI literacy, sustaining cognitive engagement, and accommodating diverse learning preferences within a rigorous pedagogical framework. 

\begin{enumerate}
    \item SocraticAI develops AI literacy through structured, reflective practice with professional-grade interaction patterns. By requiring students to articulate their reasoning, clarify problem statements, and engage in guided questioning, the system models communication behaviors used in industry settings where LLMs are increasingly embedded in development pipelines. This aligns directly with ACM/IEEE curricular guidelines emphasizing responsible and transparent use of emerging technologies as part of computational ethics and software professionalism.
    \item The system sustains cognitive engagement by enforcing a ``think–articulate–reflect'' loop. Students must explain their current understanding and attempted strategies before receiving any feedback, which reduces passive consumption and promotes active learning. Reflection prompts following each session reinforce metacognitive monitoring. In effect, SocraticAI transforms AI use from a transactional question–answer exchange into a cognitively demanding learning process.
    \item The design accommodates a range of learning styles through natural language interaction while upholding consistent pedagogical standards. Novice learners benefit from conversational scaffolds that adapt to their phrasing and conceptual level, whereas advanced students use the same system to refine code efficiency or conceptual precision. By balancing flexibility with structured constraints, the system supports both equity of access and consistency of instructional rigor across diverse cohorts.
\end{enumerate}

\begin{wrapfigure}{r}{0.4\textwidth}
 \centering
   \includegraphics[scale=0.5]{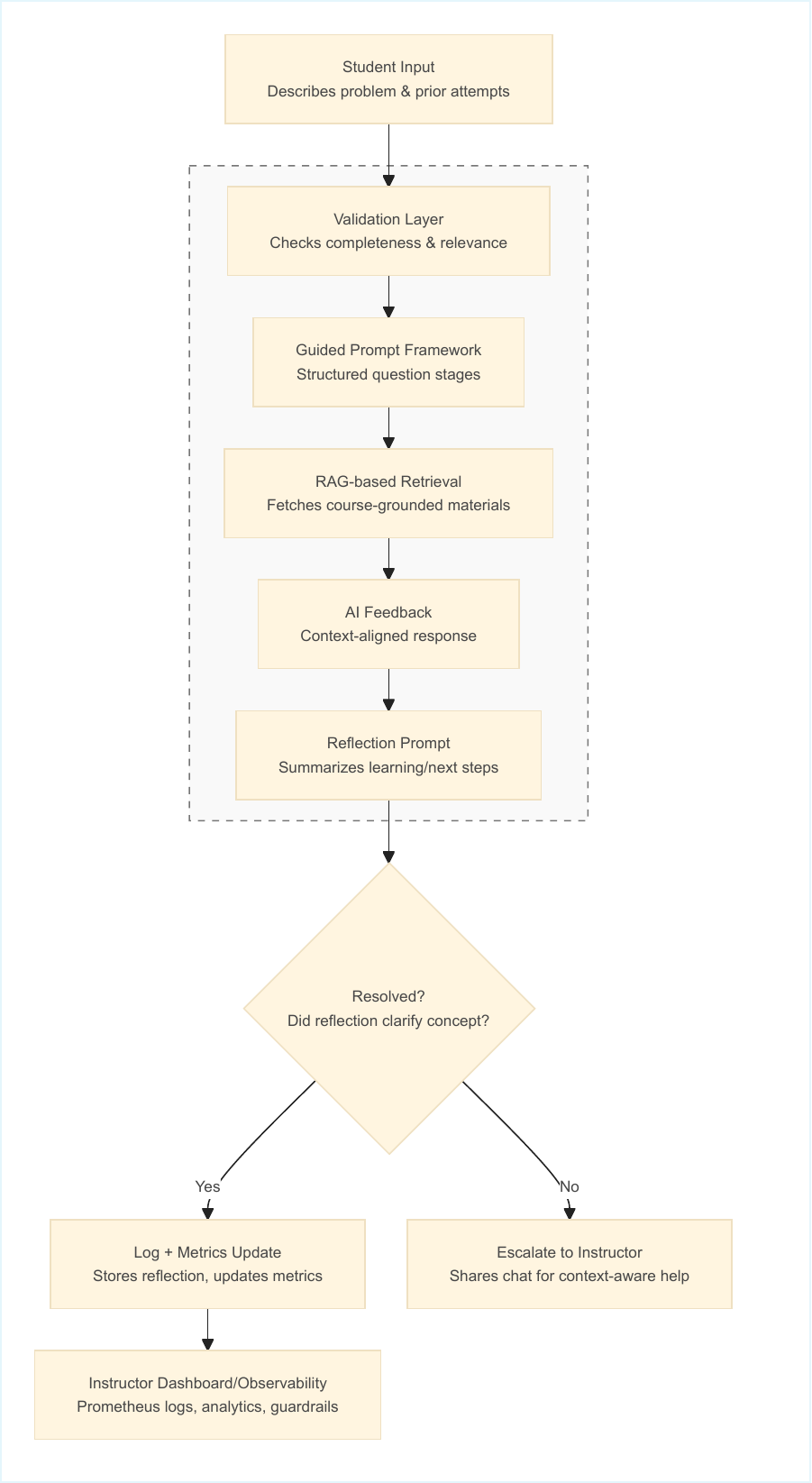}
    \caption{System Workflow Diagram}
    \label{fig:system-diagram}
     \vspace{-80pt}
\end{wrapfigure}

Taken together, these justifications position SocraticAI as a bridge between pedagogical integrity and technological innovation, teaching students not merely to \emph{use} AI, but to \emph{reason with} it.

\section{Outcomes}
Deployment in CS-1102 Introduction to Computer Science course at Ashoka University produced measurable and qualitative improvements in student learning behaviors. Over a three-week observation period, students exhibited a clear shift from vague, surface-level questions (e.g., ``My code doesn’t work'') toward precise, decomposition-oriented inquiries such as ``I implemented recursion correctly, but I’m unsure how my base case terminates.'' This linguistic evolution reflects a broader cognitive shift from debugging by trial and error to analytical problem framing.

Approximately 75\% of participants consistently provided reflective responses identifying specific misconceptions or next steps. Many reflections explicitly referenced conceptual insights (e.g., recognizing off-by-one errors as logic issues rather than syntax problems). Furthermore, instructors reported fewer repetitive, low-cognitive-load questions in office hours, suggesting that SocraticAI successfully offloaded routine guidance while preserving opportunities for deeper conceptual discussion.

Student feedback reinforced these findings. Survey responses indicated that the structured prompts ``made me slow down and think before asking for help,'' and that the reflective step ``helped me understand my own gaps.'' Several students described the system as ``training me to ask better questions,'' suggesting early evidence of growth and transferable self-regulation skills.

Controlled prompt injection experiments during the summer revealed minor vulnerabilities in the retrieval layer and context management module. These findings informed ongoing improvements in input sanitation, conversation-state handling, and adversarial testing protocols. The results underscore the importance of continuous technical iteration alongside pedagogical evaluation.

\section{Insights}
SocraticAI enhances learning quality while reducing instructor load. By requiring students to structure and reflect on their queries, the system acts as a pre-filter for instructor intervention, ensuring that escalated cases already include context, attempted reasoning, and relevant code snippets. 

Nevertheless, challenges remain. A subset of students attempted to circumvent reflective prompts or reformulate prohibited solution requests, indicating a need for adaptive constraint mechanisms and fine-tuned policy enforcement. Maintaining robust semantic grounding across diverse course materials also requires ongoing curation of the retrieval index.

\section{Conclusion}
Our findings demonstrate that structured constraints can fundamentally reframe the role of LLMs in computing education. Rather than banning or uncritically adopting generative AI, SocraticAI establishes a middle ground: a principled, scaffolded environment where students learn how to interact productively, ethically, and reflectively with AI systems.

The observed behavioral shifts indicate the potential of scaffolded AI tutors to foster not just competence, but cognitive maturity. Reflection prompts proved particularly effective in helping students internalize problem-solving techniques and in cultivating an awareness of their own learning processes. Looking forward, we envision expanding SocraticAI across more curricular contexts.

\section{Suggestions for Others}
For instructors considering similar implementations, several opportunities and pitfalls merit attention. On the opportunity side, scaffolded LLMs can scale individualized tutoring support without overwhelming teaching staff, and the required reflections generate rich data for both formative assessment and pedagogical research. Additionally, role-based dashboards provide instructors with visibility into student engagement patterns, enabling targeted interventions.

Potential pitfalls include underestimating the technical complexity of guardrails, particularly in defending against prompt injection, ensuring robust semantic retrieval, and maintaining privacy-compliant logging. Implementation also carries operational costs that must be balanced against institutional resources. To maximize effectiveness, we recommend iterative stress-testing of constraints, explicit onboarding for students to model productive interactions, and continuous monitoring of usage data to adjust limits and prompts. With these considerations, scaffolded systems like SocraticAI can be adapted to diverse institutional contexts while sustaining pedagogical integrity.
\FloatBarrier
\bibliography{references}

\end{document}